\pdfoutput=1

\documentclass[sigconf,natbib,screen]{acmart}

\usepackage{booktabs}
\usepackage{nicefrac}
\usepackage{microtype}
\usepackage{acronym}

\usepackage{tabu}
\usepackage{color}
\usepackage{algpseudocode}
\usepackage{algorithm}
\usepackage{amsbsy}
\usepackage{amsmath}
\usepackage{amsfonts}
\usepackage{bm}
\usepackage{dsfont}
\usepackage[all]{xy}
\usepackage[capitalize]{cleveref}
\usepackage[inline]{enumitem}

\usepackage{graphicx}
\usepackage{epstopdf}
\usepackage{epsfig}
\graphicspath{{figures/}}

\algrenewcommand{\algorithmiccomment}[1]{// #1}
\algnewcommand\algorithmicinput{\textbf{Input:}}
\algnewcommand\Input{\item[\algorithmicinput]}
\algnewcommand\algorithmicoutput{\textbf{Output:}}
\algnewcommand\Output{\item[\algorithmicoutput]}

\newcommand{\fedips}{\sf FedIPS}
\newcommand{\fedavg}{\sf FedAvg}
\newcommand{\fedopt}{\sf FedOpt}
\newcommand{\foltr}{\sf FOLtR}
\newcommand{\lambdalinear}{\sf \lambda Linear}
\newcommand{\cD}{\mathcal{D}}
\newcommand{\cE}{\mathcal{E}}
\newcommand{\cQ}{\mathcal{Q}}
\newcommand{\cR}{\mathcal{R}}

\newcommand{\cU}{\mathcal{U}}
\newcommand{\crqf}{\mathcal{R}_{q, f_\mathbf{w}}}
\newcommand{\fw}{f_{\mathbf{w}}}
\newcommand{\fz}{f_0}
\newcommand{\vw}{\mathbf{w}}
\newcommand{\vx}{\mathbf{x}}	

\acrodef{FL}{Federated Learning}
\acrodef{PFL}{Private Federated Learning}
\acrodef{LTR}{Learning  to Rank}
\acrodef{FLTR}{Federated LTR}
\acrodef{ULTR}{Unbiased Learning to Rank}
\acrodef{FULTR}{Federated Unbiased Learning to Rank}
\acrodef{IPS}{Inverse Propensity Score}
\acrodef{IID}[i.i.d.]{Independent Identically Distributed}
\acrodef{PBM}{ Position Based Click Model}
\acrodef{NDCG}{Normalized Discounted Cumulative Gain}
\acrodef{DP}{Differential Privacy}
\acrodef{SGD}{Stochastic Gradient Descent}

\newcommand{\mypara}[1]{\paragraph{\bf #1}}
\copyrightyear{2021}
\acmYear{2021}
\setcopyright{rightsretained}
\acmConference[ArXiv '21]{Preprint}
\acmBooktitle{ArXiv Preprint}
\acmPrice{}

 \acmDOI{} 
 
 \renewcommand\footnotetextcopyrightpermission[1]{} % removes footnote with conference information in first column
 \renewcommand\footnotetextcopyrightpermission[1]{} % removes footnote with conference information in first column

\author{Chang Li}
\orcid{0000-0002-6978-2034}
\affiliation{%
	\institution{Apple Inc.}
}
\email{chang_li3@apple.com}

\author{Hua Ouyang}
\affiliation{%
	\institution{Apple Inc.}
}
\email{hua\_ouyang@apple}

\renewcommand{\shortauthors}{Li and Ouyang}

\begin{document}
\title{Federated Unbiased Learning to Rank}

\begin{abstract}
%\ac{ULTR} studies the problem of optimizing ranking systems based on logged user interactions. 
%In a typical scenario, a user interacts with a ranking system (e.g., retrieval system), the interactions are recorded by central servers, and then \ac{ULTR} algorithms learn from these data centrally. 
%In this paper, we consider the \ac{FULTR} problem, where sending user interactions (e.g., queries, result lists and clicks) to central servers is prohibit, and the learning happens locally (e.g., on local devices).  
%Running existing \ac{ULTR} algorithms locally may not be able to solve the problem well because the amount of logged interactions in a  device is rather small. 
%To tackle this problem, we propose the $\fedips$ algorithm, which learns from user interactions on-device under the coordination of a central server and uses click propensities to address the position bias in user interactions. 
%To our best knowledge, $\fedips$ is the first \ac{FULTR} algorithm. 
%We evaluate $\fedips$ on the Yahoo and Istella datasets, and find that $\fedips$ is robust over a range of position bias.

\ac{ULTR} studies the problem of learning a ranking function based on biased user interactions. In this framework, \ac{ULTR} algorithms have to rely on a large amount of user data that are collected, stored, and aggregated by central servers. 

In this paper, we consider an on-device search setting, where users search against their personal corpora on their local devices, and the goal is to learn a ranking function from biased user interactions. Due to privacy constraints, users’ queries, personal documents, results lists, and raw interaction data will not leave their devices, and \ac{ULTR} has to be carried out via \ac{FL}.

Directly applying existing \ac{ULTR} algorithms on users’ devices could suffer from insufficient training data due to the limited amount of local interactions.
To address this problem, we propose the $\fedips$ algorithm, which learns from user interactions on-device under the coordination of a central server  and uses click propensities to remove the position bias in user interactions. 
Our evaluation of $\fedips$ on the Yahoo and Istella datasets shows that $\fedips$ is robust over a range of position biases.

\end{abstract}


\maketitle

\acresetall

%!TEX root = paper.tex

\section{Introduction}
\label{sec: introduction}

\acf{LTR} has been extensively studied in offline, online, and unbiased setups ~\citep{liu2009learning,joachims2017unbiased,li-2020-cascading,ai2021unbiased,lucchese2019learning}. In this paper, we study \ac{LTR} problems in the \acf{FL} paradigm \citep{mcmahan2016communication,mcmahan2017learning,mcmahan2021advances}, where the goal is to learn a ranking function from user interactions without centralized data collection. 
In \ac{FLTR}, a ranker is trained locally (e.g. on mobile devices or PCs), under the federation of a central server. Local devices only send essential parameters (e.g. local gradients) to the server. On-device search is one of the most important applications that motivates this setup. In this application, users search against their personal corpora on their local devices, and the goal is to learn a ranker from biased user interactions. Due to privacy constraints, users’ queries, personal documents, results lists and raw interaction data will not leave their devices. Since raw data is not centrally collected, \ac{FL} is not limited by data retention requirements, server-side storage, bandwidth and computational capacities, which enables learning with an unprecedented amount of data. 
More importantly, \ac{FL} brings new potential for protecting user privacy.
For example, with differential privacy, \ac{FL} enjoys theoretical privacy guarantees ~\citep{dwork2014algorithmic,mcmahan2017learning,geyer2017differentially,bhowmick2018protection}. Instead of tackling all these important theoretical and practical aspects of  \ac{FLTR}, 
in this paper, we focus on an important and challenging problem: federated \ac{LTR} under biased user interactions. To avoid wordiness we call this problem \emph{\acf{FULTR}}. % Need bold fonts on this!!!

There are three main challenges in \ac{FULTR}. 
Firstly, user interactions are almost always biased due to the nature of ranking~\citep{craswell2008experimental,joachims2017unbiased}. 
For example, position bias has a significant influence on user search behaviors: a document rendered at the top of a results list is more likely to be viewed and clicked than similar documents ranked at lower positions. Without addressing the position bias, a trained ranker is generally suboptimal. 
Secondly, in on-device search, the number of user interactions on each local device could be very limited comparing with traditional server-side LTR, hence directly applying existing \ac{ULTR} algorithms on each local device only leads to a suboptimal ranker. 
Finally, users only have access to their own corpora, and their behaviors are typically heterogeneous. Data distributions in each device could vary drastically~\citep{mcmahan2016communication,mcmahan2017learning}. 
This violates the typical \ac{IID} assumption in existing \ac{ULTR} algorithms.  

To tackle these challenges, we propose a simple $\fedips$ algorithm. $\fedips$ works in rounds. At the beginning of each round, a federator (central server) broadcasts an initialized ranker to participating devices. Then, local devices conduct an \ac{IPS}-weighted \ac{SGD} with their local data individually. The federator estimates the "pseudo global gradient" by combining all local gradients, and conducts a server-side \ac{SGD} to update the ranker.

We showed that $\fedips$ generates an unbiased ranker. We also extensively evaluated the performance of $\fedips$ in simulated experiments on the Yahoo\citep{chapelle2011yahoo} and Istella~\citep{Lucchese2018istella} datasets. 
These simulations mimic a real-world scenario, where all user interact with their local devices independently and leave local interactive feedback.
Our experimental results indicate that $\fedips$ is robust over a wide range of bias levels. 

%\noindent
%Our main contributions can be summarized as follows: 
%\begin{enumerate}[label=(\arabic*),leftmargin=*]
%	\item We propose the $\fedips$ algorithm to solve the position bias problem in the federated \ac{LTR}. $\fedips$ trains the ranker in a federated paradigm and uses the IPS to remove the position bias in the feedback. 
%	\item We the theoretical guarantee that $\fedips$ removes the position bias in the click feedback. 
%	\item We empirically study the performance of the proposed $\fedips$ and evaluate it in several federated unbiased LTR setups. We find that $\fedips$ is robust over a range of bias levels on both the Yahoo and Istella datasets
%\end{enumerate} 
%

%!TEX root = paper.tex

\section{Background and Related work}
\label{sec:background}
We start with the notations used in this paper. 
Let $\cQ$ be a set of queries and each $q \in \cQ$ has a set of retrieved documents $\cD_q$. 
Given query $q$, each document $d\in\cD_q$ has a binary relevance $r_{q,d} \in\{0, 1\}$. 
Let $f_{\vw}(q,d)$ be the ranking function parameterized by $\vw$. 
We denote $\crqf$ as the ranked list generated by $f_\vw$, $\crqf(k)$ as the $k$th document in $\crqf$, and $\crqf^{-1}(d)$ as the position of $d$ in $\crqf$.  

\paragraph{\bf Federated learning}
%\label{sec:federated learning}
\acf{FL} optimizes the following objective in a decentralized way~\citep{mcmahan2016communication,mcmahan2021advances}: 
\begin{equation}
\min\limits_{\vw} \ell_g(\vw) = \frac{1}{|\cU|} \sum_{u \in \cU} \frac{1}{|\cE_u|} \sum_{e \in \cE_u} \ell_l(e, f_\vw), 
\end{equation}
where $\cU$ is a subset of users, $\cE_u$ is each user's local dataset, $\ell_g(\vw) $ is the global loss function, and $\ell_l(\vw)$ is the local loss function. 
Each user interacts with local devices independently. $\fedavg$ is the earliest \ac{FL} algorithm~\citep{mcmahan2016communication}. \citet{reddi2020adaptive} proposed a more general framework $\fedopt$ with adaptive optimizers. 

%We present $\fedopt$ in~\cref{alg:fedopt} for convenience. There are two main components in $\fedopt$: $\sf ClientOpt$ and $\sf ServerOpt$. which  
%can be any optimizer, such as \ac{SGD}.  
%Each round, the server first randomly samples a subset of clients and broadcasts the global model to them. 
%Then, each client, $u \in \cU$, uses the local data and $\sf ClientOpt$  to update the global model independently in parallel, i.e.,  $\vw^*_{t, s}= \arg\min_{\vw} \frac{1}{\cE_u} \sum_{e \in \cE_u}\ell_l(e, \vw_t)$. 
% $\sf ClientOpt$  is the optimizer used by client,  such as \ac{SGD}. 
%After the local update, each client $u$ sends back local updates, $ \Delta_{t, u} = \vw^*_{t, u}-\vw_{t}$, to the server. 
%On the server side, the server estimates the global gradient, $\Delta_{t}$, by taking the average of local updates: 
%\begin{equation}
%\Delta_{t}= \frac{1}{|\cU|} \sum_{u \in\cU} \Delta_{t, u} = \frac{1}{|\cU|} \sum_{u\in\cU}  \vw^*_{t, u}-\vw_{t} .   
%\end{equation}
%Finally, the server uses $\sf ServerOpt$, the server-side optimizer, and $\Delta_{t}$ to update the global model. 

\paragraph{\bf Unbiased LTR}
%\label{sec: unbiased learning to rank}
%Counterfactual LTR is a general framework proposed by \citep{joachims2017unbiased,agarwal2019general} to tackle \ac{ULTR} in a centralized learning setup. This motivates our focus in this paper, which is the \ac{ULTR} problem in \ac{FL}.
A common approach to address position bias is to assume that user click behavior follows~\ac{PBM}~\citep{chuklin2015click} and use \acf{IPS} to compensate the difference between  true relevance and clicks~\cite{joachims2017unbiased,ai2018unbiased,wang2018position,agarwal2019general}. 
\citet{agarwal2019general} propose a general framework, which  optimizes the \ac{IPS}-weighted additive metric over clicked documents: 
\begin{equation}
\label{eq:additive metric}
	\ell(\fw | q, \cD_q) = \sum_{d\in \cD_q  \wedge c_{q,d}=1} \frac{g\left(\crqf^{-1}(d)\right)}{p_{q,d}} , 
\end{equation}
where $c_{q,d} \in \{0, 1\}$ is the click indicator. $p_{q,d}$ is the examination probability also known as propensity. $g(\cdot)$ is a position-based weighting function capturing different ranking metrics: for example DCG,  Precision@$k$, etc.~\citep{agarwal2019general,jagerman2020accelerated}.  
For simplicity, in this paper we choose $g(\crqf^{-1}(d)) = \crqf^{-1}(d)$, and our results hold for any other additive ranking metrics. 

The additive metric in \cref{eq:additive metric} is not differentiable w.r.t. $\vw$.  
In practice, we consider a surrogate loss~\citep{agarwal2019general}, which upper bounds $\cR_{q, f_\vw}^{-1}(d)$ in \cref{eq:additive metric} as follows: 
\begin{equation}
	\label{eq:surrogate metric}
	\crqf^{-1}(d) \leq 1+ \sum_{d'\in \cD_q} \max (0, 1-(\fw(q, d) - \fw(q, d'))) . 
\end{equation}
Combining  \cref{eq:additive metric,eq:surrogate metric}, we reach to the following surrogate loss function, which is used in our \ac{FULTR} setup: 
\begin{equation}
	\label{eq:surrogate loss}
	\begin{aligned}
		 \ell(\vw) &= \frac{1}{|\cQ|} \sum_{q\in\cQ} \sum_{d\in\cD_q \wedge c_{q,d}=1} \frac{h_{q, d}(\vw)}{p_{q, d}}, \\
		 \text{where }\quad	h_{q, d}(\vw)& = \sum_{d'\in \cD_q} \max (0, 1-(\fw(q, d) - \fw(q, d'))). 
	\end{aligned}
\end{equation} 
%\cref{eq:surrogate loss} is differential w.r.t. $\vw$, and can be minimized by any gradient-based optimizers, such as \ac{SGD}. 

In \ac{IPS}-based methods, the propensity score is usually assumed to be known. We make a similar assumption in this paper. 
In practice, the propensity is either estimated by controlled experiments~\citep{joachims2017unbiased,ai2018unbiased} or inferred from interaction data~\cite{wang2018position,ai2021unbiased}. 

\paragraph{\bf Federated LTR}
\ac{FL} has recently drawn attentions from the \ac{LTR} community due to growing body of research addressing privacy preservation. 
\citet{kharitonov2019federated} studies the federated online LTR and proposed $\foltr$. \citet{hartmann2019federated} uses \ac{FL} to solve the URL suggestion task for web browsing experience. 
\citet{anelli2021federank} propose $\sf FedeRank$, which is a \ac{FL} version of matrix factorization. 
Different from these methods, our focus in this paper is to deal with the position bias in  logged data.

%!TEX root = paper.tex

\section{Federated Unbiased \ac{LTR}}
\label{sec: federated unbiased LTR}
%We first formulate our learning problem in~\cref{sec: problem formulation}. 
%Then, we propose our $\fedips$ algorithm in~\cref{sec: algorithm}. 
%Finally, in~\cref{sec: discussion}, we analyze the proposed $\fedips$ and provide the theoretical guarantee on the soundness of unbiased learning. 
%
%\subsection{Problem formulation}
%\label{sec: problem formulation}

In our \ac{FULTR} setup, every client shares the same production ranking policy $\fz$, which can be synchronized during software updates.
Each user interacts with the ranking policy independently and leaves implicit feedback.
The issued queries, displayed documents, and interactive feedback are stored in local devices.
We do not make any assumption on the local data $\cQ_u$, meaning that their distributions can be heterogeneous.  
Learning in this setup has at least three challenges: 
\begin{enumerate*}
	\item The feedback is influenced by the position bias.
	\item The amount of local data $\cQ_u$ can be small. 
	\item The \ac{IID} assumption does not hold, hence existing \ac{ULTR} algorithms may not be directly applied. 
\end{enumerate*}

\subsection{$\fedips$}
\label{sec: algorithm}

We propose $\fedips$ to address these challenges. 
$\fedips$ employs $\fedopt$, a general \ac{FL} algorithm, to deal with the non-\ac{IID} challenge in \ac{FL}. 
Each client $u$ conducts the \ac{IPS}-weighted \ac{SGD} and minimizes the following objective function: 
\begin{equation}
\label{eq: local ips}
\ell_{u, IPS}(\vw_u) =  \frac{1}{|\cQ_u|}\sum_{q \in \cQ_u } \sum_{d \in \cD_q}\frac{g\left(\cR_{q, f_{\vw_{u}}}^{-1}(d)\right)}{p_{d, q}} c_{q,d} . 
\end{equation}
The goal of the global optimizer is to solve the following problem: 
\begin{equation}
\label{eq:fultr loss}
\begin{aligned}
\vw^* &= \arg\min\limits_{\vw} \ell_{g, IPS}(\vw) 
= \arg\min\limits_{\vw} \frac{1}{|\cU|} \sum_{u\in \cU} \ell_{u, IPS}(\vw),
\end{aligned}
\end{equation}
where $\ell_{g, IPS}(\vw) $ is the global objective.

Details of $\fedips$ are summarized in \cref{alg:fedips}. 
Inputs of the algorithm are: the initialized model $\vw_0$, a global learning rate $\eta_g$, and a local learning rate $\eta_l$. 
At the beginning of each round $t$, a subset $\cU$ of users are randomly drawn and the model $\vw_t$ is broadcast (Line~\ref{alg: sample users} - \ref{alg: broadcast}). 
Each client $u\in\cU$ conducts \ac{SGD} according to its local data $\cQ_u$ (Line~\ref{alg: sgd start} - \ref{alg: sgd end}). 
As the objective function in \cref{eq: local ips} is not differential, the additive metric is replaced by a surrogate metric, e.g. ~\cref{eq:surrogate metric}.
Thus, each client $u$ minimizes the following objective function: 
\begin{equation}
	\label{eq: client surrogate loss}
	\ell_{u, S}(\vw_{t, u}) = 
	 \frac{1}{|\cQ_u|}\sum_{q \in \cQ_u } \sum_{d\in\cD_q } \frac{c_{q,d} h_{q, d}(\vw_{t, u})}{p_{q,d}}, 
\end{equation}
where the surrogate loss $h_{q, d}(\vw_{t, u})$ is defined in \cref{eq:surrogate loss}.
After the local optimization, local updates $\Delta_{t, u} = \vw_{t,u}^* - \vw_{t}$, are sent to the federator (Line~\ref{alg: compute Delta}).
On the server side, the federator estimates the ``pseudo gradient''  by taking the average of all local updates, and use the server-side SGD to minimize the global loss (Line~\ref{alg: server sgd}). 
The algorithm runs for $T$ rounds and outputs the optimized ranker $\vw^{*}$. 

\subsection{Unbiased  Estimator of FULTR}
\label{sec: discussion}

In a full-information \ac{LTR} setup, we have relevance labels, and optimize the following additive metric:
 \begin{equation}
 	\label{eq:full}
 	\Psi(\fw | q, \cD_q) = \sum_{d\in \cD_q} g\left(\crqf^{-1}(d)  \right)r_{q, d}. 
 \end{equation}
In \ac{FULTR}, each client solves the problem in \cref{eq: local ips} using the feedback as ``pseudo label" instead of the relevance ground truth.
In this section, we show that the metric in \cref{eq: local ips} is an unbiased estimator of the full-information metric in~\cref{eq:full}. 
The proof methodology is inspired by \citep{agarwal2019general}. 
First, we denote $e_d \in \{0, 1\}$ as the examination indicator of a document $d$. 
$e_d=1$ means $d$ is examined by the user, otherwise it is $0$. 

Given a user $u$, we have:  
\begin{equation}
	\begin{aligned}
		&\mathbb{E}_{e}\left[ \frac{1}{|\cQ_u|}\sum_{q \in \cQ_u } \sum_{d\in\cD_q}\frac{g(d) c_{q,d}}{p_{q,d}}    \right] 
		\stackrel{(a)}{=} 
		  \frac{1}{|\cQ_u|} \sum_{q \in \cQ_u }  \sum_{d\in \cD_q}\mathbb{E}_{e}\left[\frac{g(d) e_d r_{q, d}  }{p_{q,d}}  \right] \\  
		&= 
		 \frac{1}{|\cQ_u|} \sum_{q \in \cQ_u }  \sum_{d\in \cD_q} \frac{ g(d) p(e_d=1) r_{q, d}  }{p_{q,d}}  \\  
		&\stackrel{(b)}{=} 
		 \frac{1}{|\cQ_u|} \sum_{q \in \cQ_u }	\sum_{d\in \cD_q} g(d) r_{q, d}  
		= 
		 \frac{1}{|\cQ_u|} \sum_{q \in \cQ_u } \Psi(f_w| q, \cD_q), 
	\end{aligned}
\end{equation}
where $g(d)$ is a shorthand for $g( \crqf^{-1}(d))$. 
Equation (a) comes from the \ac{PBM}~\citep{chuklin2015click}: where $c_{q,d}=e_{d} \cdot r_{q,d}$. 
Equation (b) holds because $p(e_d=1) = p_{q,d}$. 
This shows that each client conducts an unbiased learning. 

Similarly we show that on the server side the global objective function $\ell_{g, IPS}(\vw)$ in \cref{eq:fultr loss} is also unbiased: 
\begin{equation}
	\begin{aligned}
		&\mathbb{E}_e[\ell_{g, IPS}(\vw)] 
		=  \mathbb{E}_e\left[ 
		\frac{1}{|\cU|} \sum_{u \in \cU} \frac{1}{|\cQ_u|} \sum_{q\in\cQ_u} \sum_{d\in \cD_q} \frac{g(d) c_{q, d}}{p_{q,d}} 
		\right] \\
		&=   
		\frac{1}{|\cU|} \sum_{u \in \cU}\frac{1}{|\cQ_u|} \sum_{q\in\cQ_u} \sum_{d\in \cD_q} \mathbb{E}_e\left[\frac{1}{p_{q,d}} \cR_{q}^{-1}(d)
		\right] \\
		&= 	\frac{1}{|\cU|} \sum_{u \in \cU}\frac{1}{|\cQ_u|} \sum_{q\in\cQ_u}   \Psi(f_w| q, \cD_q).
	\end{aligned}
\end{equation}

\begin{algorithm}[tb]
	\caption{$\fedips$}
	\label{alg:fedips}
	\begin{algorithmic}[1]
		\Input  $\mathbf{w}_0$, $\eta_g$ and $\eta_l$ 
		\Output $\vw^*$ 
		\For{$t=0, 1,  \ldots T-1$}
		\State\label{alg: sample users} Sample a client subset $\cU$ 
		\State\label{alg: broadcast} $\vw_{t,u} \leftarrow \vw_{t}, \forall u \in \cU$ \Comment{Broadcast $\vw_{t}$}
		\For{$u \in \mathcal{U}$} \Comment{In parallel}
		\For{$ d_i \in \cD_q$ and $q \in \cQ_u$}  \label{alg: sgd start} 
		\State $g_u^{i} \leftarrow  \frac{1}{p_{d_i}}  \nabla 	h_{q_i, d_i}(\vw_{t, u})$	 \Comment{Client gradient}
		\State $\mathbf{w}_{t, u} \leftarrow \mathbf{w}_{t, u} - \eta_l g_u^{i}$  \Comment{Client update}
		\EndFor \label{alg: sgd end}
		\State\label{alg: compute Delta} $\Delta_{t, u} \leftarrow \mathbf{w}_{t, u}-\mathbf{w}_t$ 
		\EndFor
		\State $\mathbf{w}_{t+1} \leftarrow \mathbf{w}_{t} + \eta_g \frac{1}{|\cU|} \sum_{u \in\mathcal{U}} \Delta_{t, u}$ \label{alg: server sgd} \Comment{Sever update}
		\EndFor
		\State $\vw^* \leftarrow \vw_{T}$
	\end{algorithmic}
\end{algorithm}

%!TEX root = paper.tex

\section{Experimental Setup}
\label{sec: experimental setup}

\mypara{Dataset} 
Using offline collected datasets and simulated experiments to evaluate \ac{FL} and \ac{ULTR} algorithms is a widely adopted practice ~\citep{mcmahan2016communication,reddi2020adaptive,joachims2017unbiased,jagerman2020accelerated}. 
In our experiments, we choose two public available \ac{LTR} datasets: Yahoo~\citep{chapelle2011yahoo} and Istella~\citep{Lucchese2018istella}. 
These datasets contain large collections of queries $\cQ$ from real-world search engines. 
Each query $q$ is attached with a set of candidate documents $\cD_q$.
Each query-document pair is represented by a feature vector $\vx_{q,d}$.
%: a $519$-dimensional vector for the Yahoo dataset and a $220$-dimensional vector for the Istella one, and graded into a $5$-scale relevance score. 
For pre-processing, we filter out queries where all documents have the same score. 
After pre-processing, the Yahoo dataset contains $14,377$ queries with $32.51$ documents per query on average. The Istella dataset contains $32,625$ queries with $103.73$ documents per query on average. 
We conduct the query-level normalization, and scale features to the range of $[0, 1]$. 

\mypara{Logging policy}
The logging policy $\fz$ mimics the production ranker in a real-world search engine. 
We follow a commonly used setup~\citep{joachims2017unbiased,jagerman2020accelerated} and sample $1\%$ queries to train a linear ranker in a full-information manner.

\mypara{Click simulator}
We follow the \ac{PBM} designed by~\citep{jagerman2020accelerated} and simulate clicks as follows: 
\begin{equation}
p(c_{q, d} = 1) = 
\begin{cases}
p(e_{q,d, u}), 	&\text{if } r_{q, d} \in \{3, 4\} \\
0.1p(e_{q,d, u}),              & \text{otherwise}
\end{cases}
\end{equation}
where $p(e_{q,d, u})$ is the user-dependent examination probability. 
This is to mimic the real-world scenario where every user's position bias is different. 
Give user $u$,  we define $p(e_{q,d, u})  =(\frac{1}{\cR_{q, \fz}^{-1}(d)})^{\gamma_s}$, 
where $\gamma_s \geq 0$ is a user-dependent position bias factor. 
In our experiments, we sample $\gamma_s$ from a left-truncated Gaussian distribution $\mathcal{N}(\gamma, 0.1)$.  
%$\gamma$ is the average bias parameter over all users.
%By choosing different $\gamma$, we can define different levels of position bias.  
%The larger $\gamma$, the more server position bias, and, when $\gamma = 0$, there is no position bias. 
%Other than further specified, we use $\gamma=1.0$ as the default setup.

\mypara{Federated learning setup}
We simulate $|\cU|$ clients, each of which issues $5$ queries per round.~\footnote{We choose $5$ because users may have limited interactions with devices. In our experiments, we find that the number of clicks is more important than the number of queries.}
On all devices, $K$ documents are displayed for each query. 
During each global round, a user clicks $m$ documents. 
To study the impact of position bias, we choose $|\cU|=2,000$, $K=5$ and $m=10$. 

\mypara{Baselines}
To our best knowledge, $\fedips$ is the first \ac{FULTR} algorithm.
So we design two non-FULTR baselines for comparison. 
The first one is the vanilla $\fedavg$ algorithm~\citep{mcmahan2016communication}. 
The second baseline is the linear LambdaRank~\cite{burges2010ranknet} trained in a full-information manner. 
We name it  $\lambdalinear$.  
We follow the setup in \citep{joachims2017unbiased,jagerman2020accelerated} and use the linear ranker:
\begin{equation}
	f_{\vw}(q, d) = \vw^T\vx_{q,d}.
\end{equation}
We believe that the proposed $\fedips$ is general enough to accommodate more sophisticated models, such as neural networks~\citep{pasumarthi2019tf}. 

For $\fedips$ and $\fedavg$, we use a grid search to choose parameters from ranges of: 
$\eta_l \in \{0.00001, 0.0001, \ldots, 0.1 \}$ and 
$\eta_g \in \{0.05, 0.5, 1, 2\}$.
For $\lambdalinear$, we choose the learning rate from \break$\{0.001, 0.005, 0.01, 0.03, \ldots, 0.09, 0.1 , 0.3, \ldots, 0.9 \}$. 

NDCG@5 is used as the evaluation metric.

\begin{figure*}[t]
	\centering
	\includegraphics{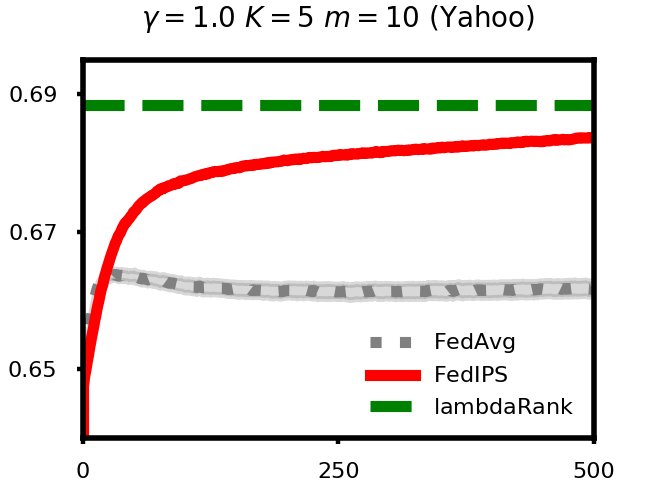}
	\includegraphics{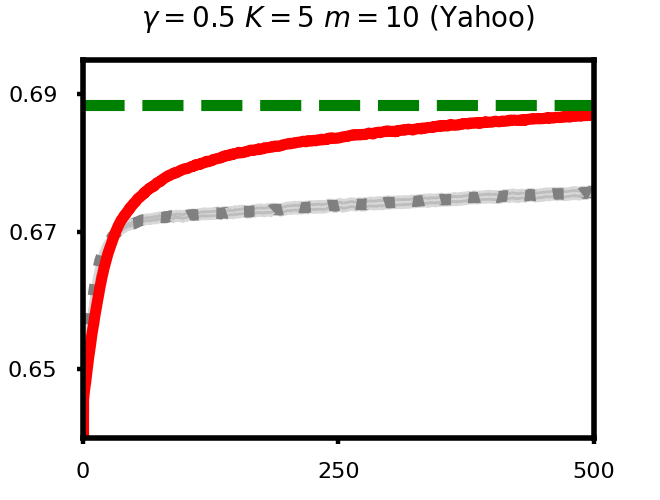}
	\includegraphics{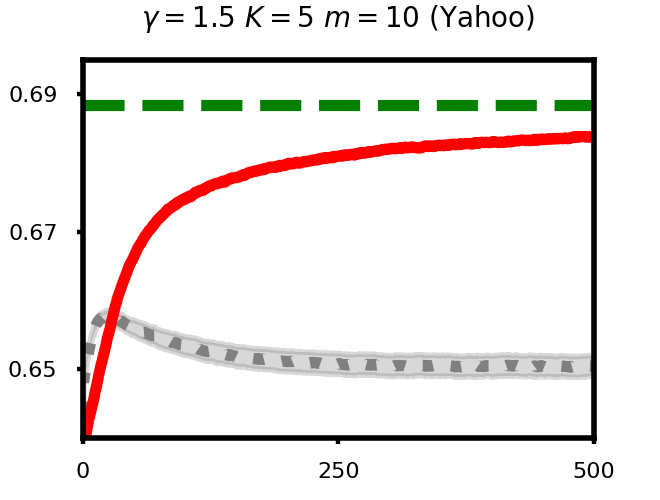}
	\includegraphics{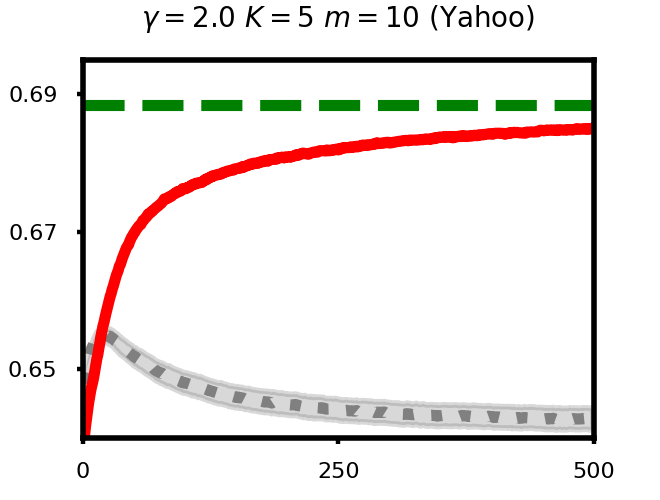}	
	\\
	\includegraphics{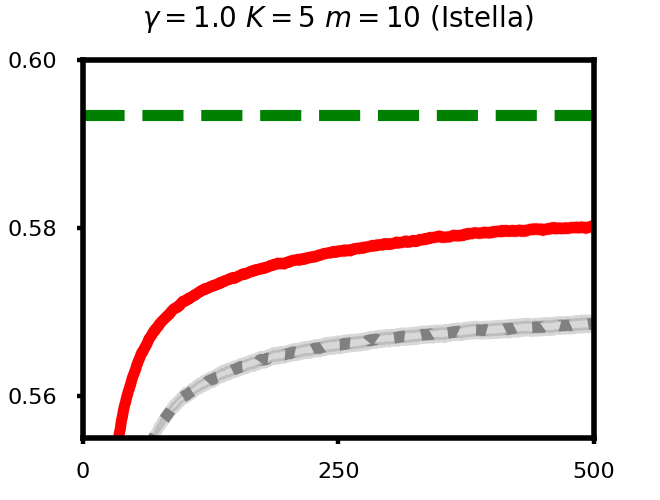}
	\includegraphics{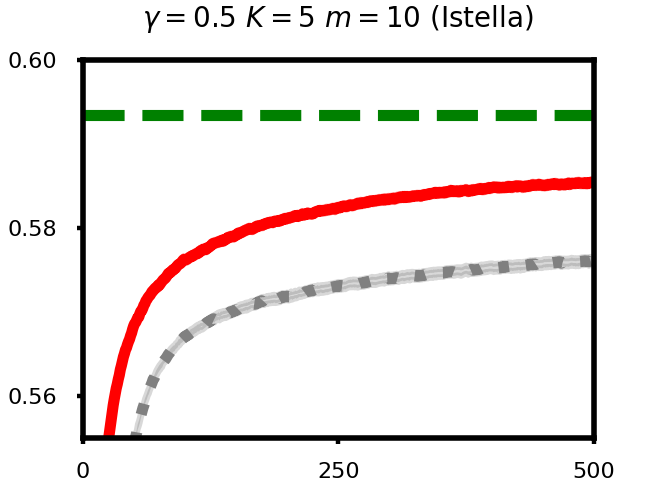}
	\includegraphics{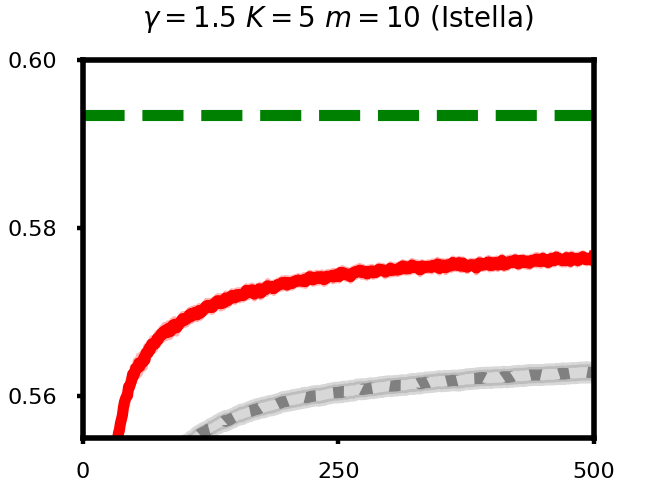}
	\includegraphics{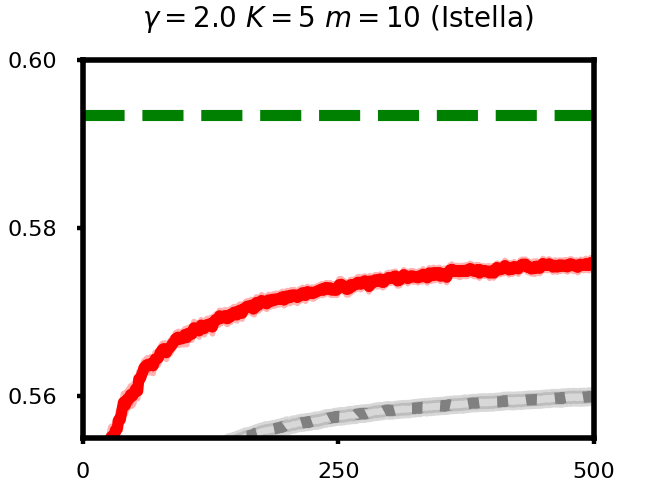}
	\caption{Compare NDCG@$5$ for $\fedips$, $\fedavg$ and $\lambdalinear$. X-axis is the number of rounds. $K$ is the number of positions and $m$ is the number of clicks per user. Results are averaged over $30$ repeats. Shaded areas indicate standard error.}
	\label{fig: sigir}
	
\end{figure*}

\section{Experimental results}

We first compare $\fedips$ with non-FULTR baselines. As demonstrated in the left column of \cref{fig: sigir} where $\gamma=1$ is picked as an example, on both Istella and Yahoo datasets, $\fedips$ significantly outperforms $\fedavg$. In Yahoo dataset, $\fedavg$ starts to taper off after only a few iterations, while $\fedips$'s NDCG keep growing even after $500$ iterations, almost approaching the full-informational $\lambdalinear$.  

We then study the impact of varying the level of position bias $\gamma$ and results are shown in \cref{fig: sigir}.
We run $\fedips$ and $\fedavg$ with $\gamma \in \{0.5, 1.0, 1.5, 2.0\}$. We observe that $\fedips$ is fairly robust to the different degrees of position biases. Although $\gamma=0.5$ gives the best NDCG, their differences are small. On the other hand, we notice a huge drop in the performance of $\fedavg$ when $\gamma$ becomes large. 
This indicates that our $\fedips$ handles position bias pretty well.  

\begin{figure}
	\centering
	\includegraphics{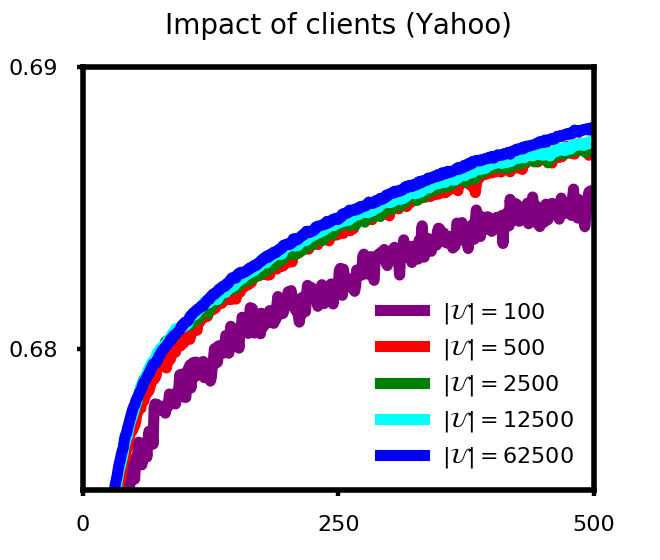}
	\includegraphics{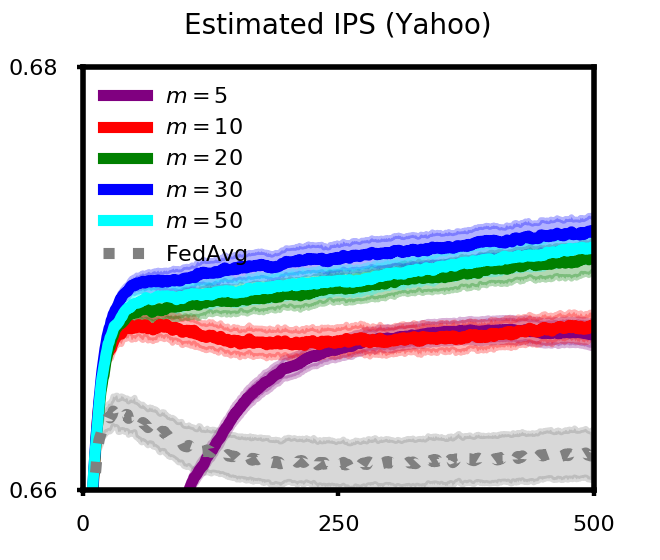}
	\caption{ 
		Left: impact of the numbers of clients $|\cU|$ on $\fedips$.
		Right: estimated \ac{IPS} with different number of user clicks $m$. 
		X-axes are numbers of rounds and Y-axes are NDCG@$5$. }
	\label{fig: additional}
\end{figure}

We also study the impact of different numbers of clients (local devices). 
We expect better performance of $\fedips$ with more clients registered for FL. 
We simulated $|\cU| \in \{100, 500, 2500, 12500, 62500 \}$. 
Results are reported in the \cref{fig: additional} (Left). \footnote{Due to limited space, in the following experiments, we only report results on the Yahoo datasets, but similar results are observed on the Istella dataset.} 
We observe that when $|\cU|$ reaches a certain level, e.g., $|\cU| = 2500$, increasing $|\cU|$ only marginally improves the metrics of $\fedips$. Specifically, $\fedips$ with $|\cU|=100$ falls behind others and the learning curves of $|\cU| \in \{100, 500\}$ have large fluctuations since the estimated global gradient is noisy with small $|\cU|$. 
When $|\cU| \geq 2500$, $\fedips$ exhibits similar convergence trends. 
Our speculation is that when a sufficient amount of clients are involved, $\fedips$ estimates a rather accurate global gradient, and the variance in the estimator is no longer a bottleneck.

Finally, we experiment with a simple way to estimate \ac{IPS} in \ac{FULTR} without intervention. 
We extend the regression-based EM approach by~\citet{wang2018position} to \ac{FULTR}. 
Briefly, a function $F$ is used to estimate the hidden relevance $r_{q,d}$, which is then used in the M-step to infer the propensity score~\citep{wang2018position}.  
In our experiments, we choose $F$ as a linear function and use $\fedopt$ to optimize it. 
The updated $F$ is used to estimated personalized \ac{IPS}, i.e., running a similar version of Algorithm~1 in~\citep{wang2018position} on local client.

We conduct experiments with different numbers of clicks $m$, and report the results in \cref{fig: additional} (Right).  
We observe that for all $m$ values, $\fedips$ with the estimated \ac{IPS} eventually outperforms $\fedavg$. 
In the most sparse setting where users only provide $m=5$ clicks, $\fedips$ outperforms $\fedavg$ only after about $100$ rounds. 
We also observe performance gaps between the estimated IPS and real IPS (in contrast to \cref{fig: sigir}). 
This indicates that although our estimated \ac{IPS} is helpful in addressing the position bias, here are still a lot of room for improvement.

%\input{6-related work.tex}
%!TEX root = paper.tex

\section{Conclusion}
\label{sec:conclusion}

In this paper, we study the \acf{FULTR} problem, an important challenge in federated LTR. 
We propose the $\fedips$ algorithm as a solution. 
We conduct extensive experiments on public datasets to evaluate $\fedips$, and show that $\fedips$ outperforms the biased $\fedavg$, and approaches the performance of the full-information $\lambdalinear$.

We point out a few directions for future research.
First, to enjoy theoretical guarantees on privacy preservation, it is a common practice to apply randomized mechanisms to FL algorithms, as studied by the differential privacy~\citep{dwork2014algorithmic,mcmahan2017learning} community. 
Another challenging problem is to better estimate propensities in the FL setup. This will help bridge the gap that we observed in our last set of experiments. Evaluating $\fedips$ with sophisticated models such as nonlinear neural networks or boosting trees will also be very interesting.

%\begin{acks}
%Thank you all! 
%\end{acks}

\bibliographystyle{ACM-Reference-Format}
\bibliography{paper}

%\clearpage
%\appendix
%\input{appendix}

\end{document}